# X-Shooting ULLYSES: Massive Stars at Low Metallicity

Jorick S. Vink[1]
Paul Crowther[2]
Alex Fullerton[3]
Miriam Garcia[4]
Fabrice Martins[5]
Nidia Morrell[6]
Lida Oskinova[7]
Nicole St Louis[8]
Asif ud-Doula[9]
Andreas Sander[10]
Hugues Sana[11]
Jean-Claude Bouret[12]
Brankica Kubatova[13]
Pablo Marchant[11]
Lucimara P. Martins[14]
Aida Wofford[15]
Jacco van Loon[16]
O. Grace Telford[17,18]
Ylva Götberg[18]
Dominic Bowman[11,19]
Christi Erba[20]
Venu Kalari[21]
and The XShootU Collaboration

[1] Armagh Observatory and Planetarium, UK
[2] Department of Physics & Astronomy, University of Sheffield, UK
[3] Space Telescope Science Institute, Baltimore, USA
[4] Centre for Astrobiology (CSIC-INTA), Torrejón de Ardoz, Madrid, Spain
[5] Montpellier Universe and Particles Laboratory, Montpellier University, France
[6] Las Campanas Observatory, Carnegie Observatories, Chile
[7] Institute for Physics and Astronomy, University of Potsdam, Germany
[8] Department of Physics, University of Montreal, Canada
[9] Penn State Scranton, Dunmore, PA, USA
[10] Astronomy Centre, Heidelberg University, Germany
[11] Institute of Astronomy, KU Leuven, Belgium
[12] Aix Marseille University, CNRS, CNES, LAM, Marseille, France
[13] Astronomical Institute of the Czech Academy of Sciences, Ondřejov, Czech Republic
[14] NAT – São Paulo City University, Brazil
[15] Institute of Astronomy, National Autonomous University of Mexico, Ensenada, Mexico
[16] Lennard-Jones Laboratories, Keele University, UK
[17] Department of Astrophysical Sciences, Princeton University, USA
[18] The Observatories of the Carnegie Institution for Science, Pasadena, USA
[19] School of Mathematics, Statistics and Physics, Newcastle University, UK
[20] Physics and Astronomy, East Tennessee State University, USA
[21] Gemini Observatory/NSF's NOIRLab, La Serena, Chile

**The Hubble Space Telescope has devoted 500 orbits to observing 250 massive stars with low metallicity in the ultraviolet (UV) range within the framework of the ULLYSES program. The X-Shooting ULLYSES (XShootU) project enhances the legacy value of this UV dataset by providing high-quality optical and near-infrared spectra, which are acquired using the wide-wavelength-coverage X-shooter spectrograph at ESO's Very Large Telescope. XShootU emphasises the importance of combining UV with optical spectra for the consistent determination of key stellar parameters such as effective temperature, surface gravity, luminosity, abundances, and wind characteristics including mass-loss rates as a function of metallicity. Since uncertainties in these parameters have implications across various branches of astrophysics, the data and modelling generated by the XShootU project are poised to significantly advance our understanding of massive stars at low metallicity. This is particularly crucial for confidently interpreting James Webb Space Telescope (JWST) data of the earliest stellar generations, making XShootU a unique resource for comprehending individual spectra of low-metallicity stars.**

## The role of metallicity

Over the past few decades, it has become clear that metallicity — the relative amount of heavy elements like iron (Fe) — significantly influences the fundamental properties and behaviour of stars. This chemical make-up affects crucial stellar characteristics such as how hot they are and how they pulsate, while in the case of massive stars it also sets the rate of mass loss from their powerful stellar winds, as hot stars are driven by radiation pressure on metallic line opacity. Since the beginning of the Universe, metallicity has been on the rise. This increase is due to the chemical enrichment caused by stellar winds and supernova (SN) explosions. Ultimately, this has led to the metal-rich environment we find in our Milky Way galaxy today, approximately corresponding to the metallicity of our Sun.

But here's the exciting part: it is not just redshift and cosmic time that determine a galaxy's metal content. Another key factor is a galaxy's mass. This means we have a unique opportunity to explore the more pristine conditions similar to the early Universe by studying low-metallicity dwarf galaxies right in our cosmic backyard. In particular, the Small and Large Magellanic Clouds (SMC and LMC), which contain just about 20% and 50% of the heavy elements found in our Sun are ideal laboratories. Exploring these 'metal-poor' galaxies offers us a glimpse of what the Universe was like in its infancy. It is like looking at a cosmic time capsule, telling us the story of the evolving Universe through the lens of stellar metallicity.

## The metallicity-dependent fate of massive stars

The Universe still holds many secrets. Of particular relevance to our objectives are the processes governing the formation of black holes (BHs) over time and as a function of metallicity, as well as the physics underlying the occurrence of superluminous supernovae (SLSNe) in low-metallicity environments. Among these mysterious explosions is a subset that might involve an extremely disruptive phenomenon known as a pair-instability SN, where the entire star is obliterated, and which was initially theorised back in the 1960s (Fowler & Hoyle, 1964). However, concrete observations of such events in the real world have so far remained elusive.

Understanding the boundary between BH and pair-instability SNe is paramount for our understanding of gravitational wave (GW) events. While GW events represent just a small fraction of the possible outcomes of stellar evolution, the broader question of whether a massive star ultimately becomes a neutron star, a BH, or experiences a pair-instability SN is of profound significance in our understanding of how the Universe becomes enriched with elements. A single pair-instability SN explosion resulting from a 300-solar-mass progenitor star could contribute more heavy elements than the entire range of stars from a fully sampled initial mass function below it (Langer, 2009).

The ubiquitous property of the most massive stars involves their metallicity-dependent mass loss, which is the key physical process that sets the boundary between BH formation and pair instability (for example, Yusof et al., 2013; Köhler et al., 2015). For this and many other reasons we need to test our theoretical predictions of metallicity-dependent winds against large sets of reliable empirical data (Vink et al., 2023, XShootU I).

## The powerful combination of ultraviolet and optical spectroscopy

The ultraviolet (UV) region of the spectrum provides access to a unique suite of stellar wind diagnostics, in particular the P Cygni lines associated with abundant chemical species like C IV (Figure 1). The blue boundary of the P Cygni profile offers a reliable means to gauge the terminal wind velocity (Prinja, Barlow & Howarth, 1990; Hawcroft et al., 2023, XShootU III). However, on its own the UV part of the spectrum presents certain challenges, stemming from uncertainties in the ionisation state which limit the ability to quantify the mass-loss rate in massive stars (Lamers & Leitherer, 1993). The optical range emerges as a critical complement to UV spectra in order to acquire accurate stellar and wind parameters (as depicted in Figure 1; Fullerton et al., 2006; Oskinova, Hamann & Feldmeier, 2007; Puls, Vink & Najarro, 2008).

It is not just the wide wavelength range — from the UV to the optical — that is crucial for building an appropriate framework of metallicity-dependent stellar winds, but also the size of the sample. Historically, most investigations into the winds of massive stars featured limited sample sizes (typically around 10), along with varying coverage across different instruments, wavelength ranges, and analysis tools.

## The HST ULLYSES opportunity

Developments in sample size are currently undergoing a revolution, thanks to the HST ULLYSES project. This initiative, a Director's Discretionary Time (DDT) endeavour of 1000 orbits, was conceived by a dedicated working group. The project comprises two distinct parts: a 500-orbit programme focused on young T Tauri stars, and a comparable allocation of HST orbits for the massive star component. The latter aspect of the DDT project received substantial support from the massive star community, including backing from entities like the IAU Commission G2 on Massive Stars. The project encompasses the entire range of spectral types and luminosity classes among OB-type stars within the LMC and SMC regions. This unprecedented coverage includes representatives of all stars with masses exceeding 10 $M_{\odot}$, as indicated in the Hertzsprung–Russell diagrams of the LMC and SMC presented in Figure 2. XShootU also includes a small number of stars in even lower-redshift galaxies.

The dataset generated by this project will leave a lasting legacy, not only enhancing our understanding of stellar winds but also serving as a crucial spectral library from which to construct accurate population synthesis models. Recognising the pivotal role played by the optical wavelength range in determining both stellar and wind parameters, the community strongly advocated that this exceptional opportunity presented by a substantial UV legacy dataset should be complemented by an equally substantial and high-quality dataset in the optical. Hence

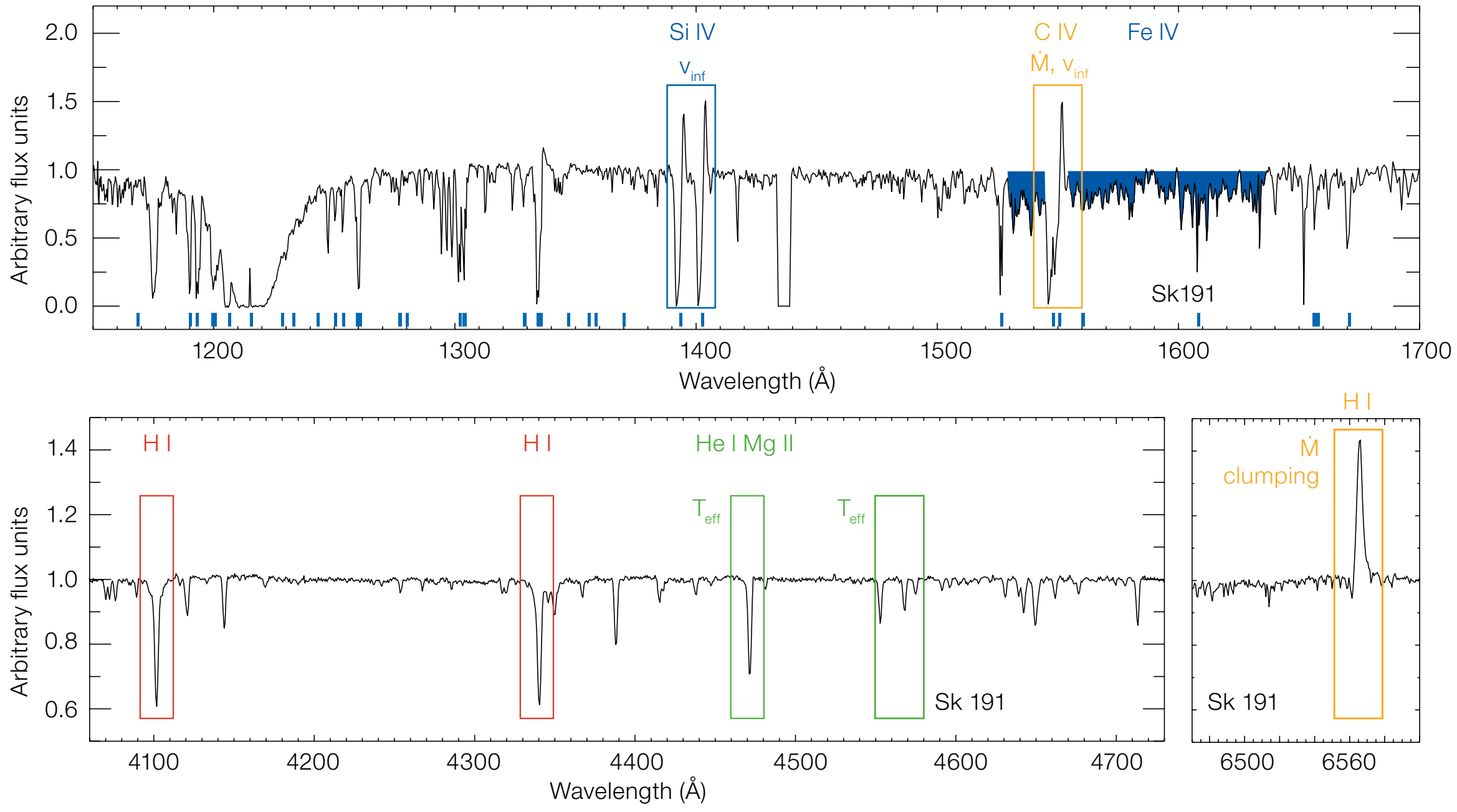

**Figure 1.** UV (ULLYSES) and optical (XshootU) spectroscopy of the early B supergiant Sk 191 in the SMC, highlighting key photospheric (red and green) and wind (blue, orange) diagnostics.

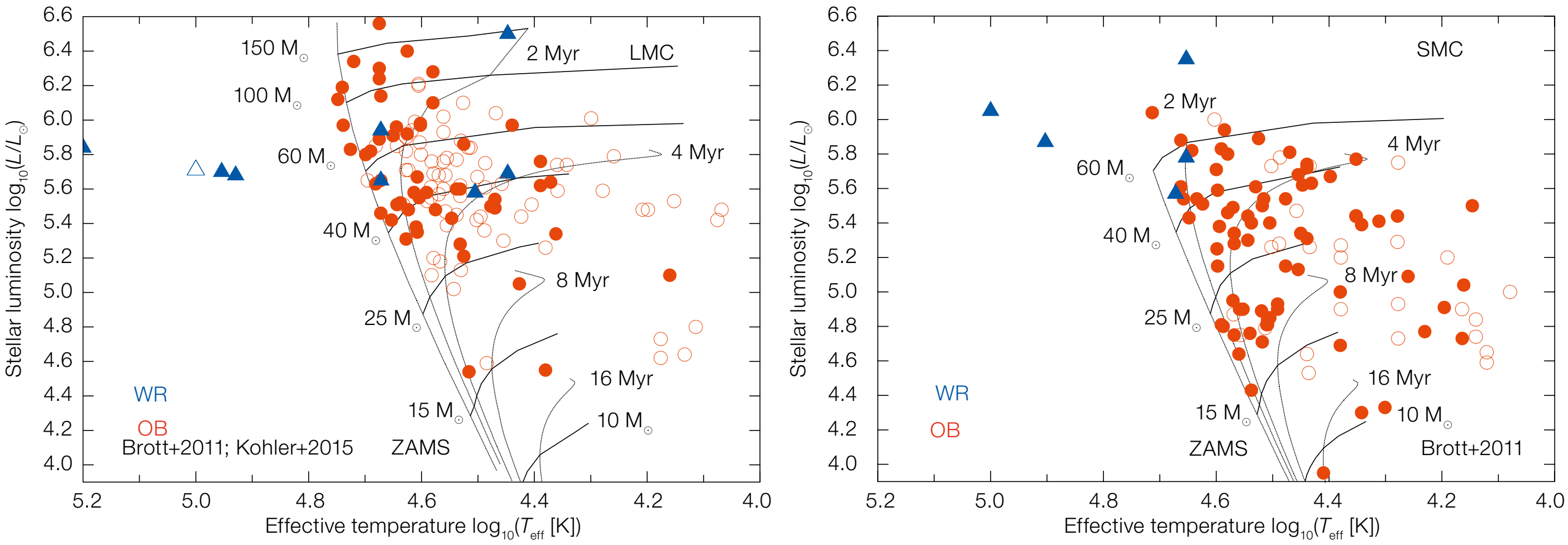

**Figure 2.** Hertzsprung–Russell diagrams highlighting the location of ULLYSES/XshootU targets (OB: red circles, WR: blue triangles) in the LMC (left) and SMC (right), compared to evolutionary models from Brott et al. (2011) and Köhler et al. (2015).

the inception of the XShooting ULLYSES (XShootU) project[1].

As a bonus, the X-shooter spectrograph offers the additional prospect of establishing a spectral library in the near-infrared. Given that many of the world's newest telescopes have their prime capabilities in this part of the electromagnetic spectrum, it is important that we extend our UV+ optical analysis tools into this regime.

## The XShootU advanced data products

Providing advanced data products to the community is one of the main goals of the XShootU collaboration. The optical (UVB+VIS) data have already been reduced and an example of the UVB data is shown in Figure 3. A comprehensive account of these higher-level data products is presented in XShootU's Data Release 1 (Sana et al., 2024, XShootU II), with the data reduction processes for the near-infrared spectra following later.

The raw data first underwent data reduction using the ESO X-shooter pipeline, with a focus on determining response curves. This involved ensuring equal flat-fielding for the scientific targets and flux standard stars, improving the flux standard models. We then processed the pipeline products using our own procedures, generating a range of advanced data products. These included corrections for slit losses, absolute flux calibration, (semi-)automatic rectification to the continuum, and removal of telluric lines. Additionally, the spectra from different epochs were corrected for barycentric motion and combined to create a single, flux-calibrated spectrum covering the entire optical range with the highest possible signal-to-noise ratio.

Our analysis revealed an undocumented recurring ghost artefact present in the raw data. We further introduced an enhanced flat-fielding strategy to minimise artefacts when scientific targets and flux standard stars were observed on different nights. The improved flux standard models and a new set of reference points allowed us to significantly reduce artefacts in the correction of response curves, especially in the wings of the Balmer lines, where discrepancies decreased from a few percent of the continuum level to less than 0.5 percent.

Furthermore, we confirmed the existence of a radial velocity shift of approximately 3.5 km $s^{-1}$ between the UVB and VIS arms of X-shooter and demonstrated the absence of short-term variations affecting radial velocity measurements. We achieved a radial velocity precision of less than 1 km $s^{-1}$ on sharp telluric lines and between 2 and 3 km $s^{-1}$ on data with the highest signal-to-noise ratios.

This post-processing provided three data products for each target: (i) 2D spectra for each exposure before and after instrument response correction; (ii) 1D spectra as initially generated by the X-shooter pipeline, followed by response correction and various processing steps, including absolute flux calibration, telluric line removal, normalisation, and barycentric correction; and (iii) co-added, flux-calibrated, and rectified spectra spanning the full optical range, combining all available XShootU exposures. For the majority of targets, the final signal-to-noise ratio per resolution element exceeds 200 in both UVB and VIS co-added spectra.

Reduced data and the most important advanced data products will be made accessible to the scientific community via the ESO Science Archive Facility in line with ESO's policy for Large Programmes. Together with the HST UV ULLYSES data available for MAST2, they enable a variety of scientific investigations, ranging from detailed studies of stellar atmospheres and stellar winds to the creation of empirical libraries for population synthesis.

## The XShootU project

XShootU is a community-focused project, where collaboration is organised through 14 working groups that are open for participation to any scientist. We already held a number of online and in-person meetings to facilitate discussions and collaboration among researchers engaged in the spectroscopic analysis of the XShootU data-sets. The topics range from determining the stellar and wind properties of the sample on a broader scale to addressing specialised issues such as the impact of rotational mixing through abundance studies. The primary spectroscopic analysis tools employed are non-LTE codes that include a stellar outflow in spherical geometry, including CMFGEN (Hillier & Miller, 1998), FASTWIND (Puls et al., 2020), and PoWR (Gräfener et al., 2002), or the plane parallel non-LTE code TLUSTY (Hubeny & Lanz, 1995) for stars lacking strong winds.

Research has demonstrated that the characteristics of structured winds significantly influence empirical calculations of mass-loss rates, thereby shaping our comprehension of stellar evolution. Nevertheless, the extent to which mass loss is influenced by wind clumping remains an open question. Within XShootU, we investigate the impact of clumping on spectral and luminosity classes at various metallicities. The spectral modelling leverages advanced model atmosphere codes, capable of addressing clumping properties with varying degrees of complexity (Sander et al., in preparation).

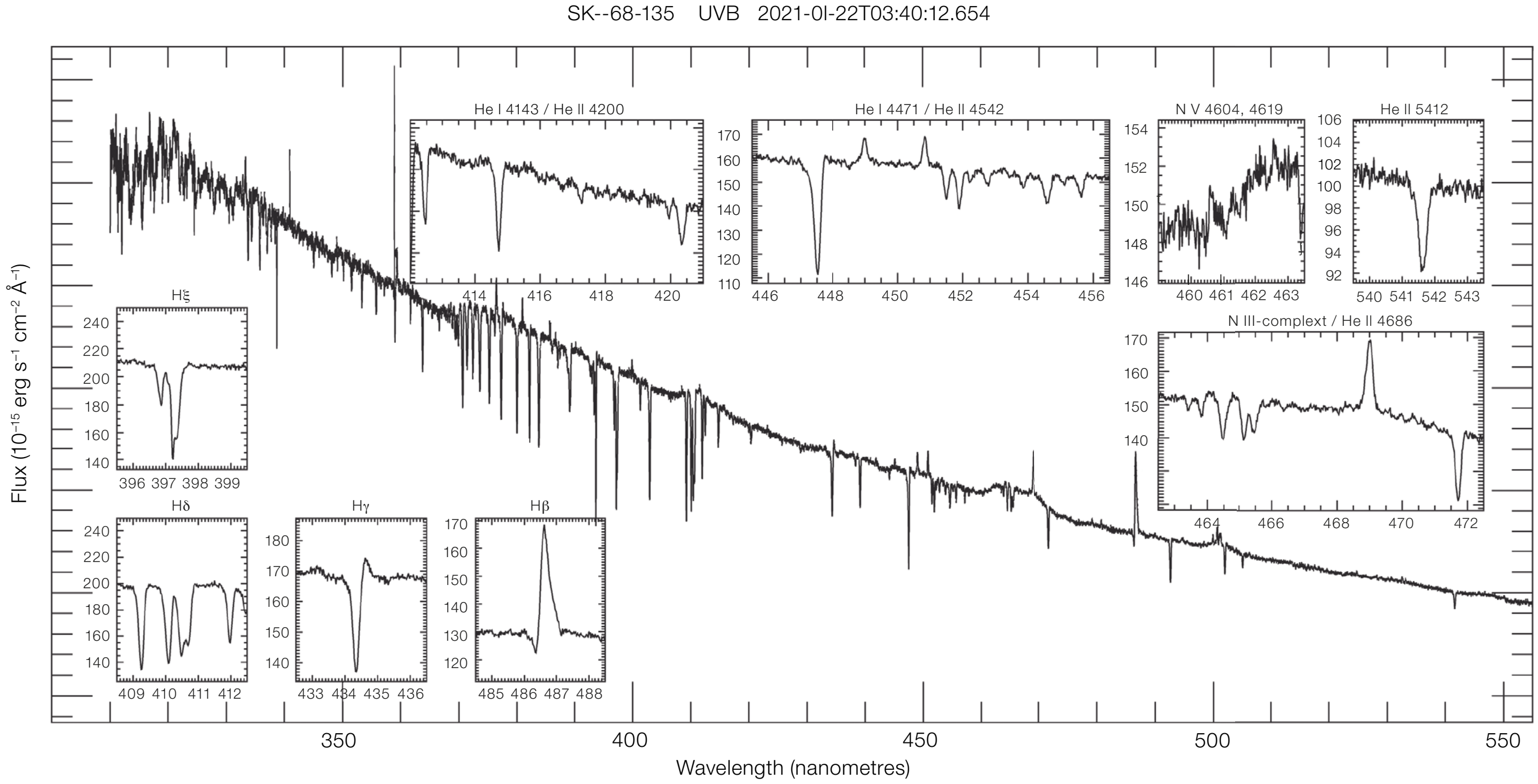

Figure 3. VLT/Xshooter UVB spectroscopy of the late O supergiant Sk –68° 135 in the LMC.

## Early spectral synthesis results

Some of the key science goals from XShootU concern the role of metallicity in setting the mass-loss rate from massive stars and how metallicity may affect internal mixing. In order to achieve these objectives we require the determination of mass-loss rates as well as abundances in both the UV and the optical.

An early study into interior mixing was performed by Martins et al. (submitted to A&A) using the CMFGEN code. Figure 4 shows nitrogen (N) abundances versus the surface gravity (g) of O6.5–O9 dwarfs in the different environments of the SMC, the LMC, and the Milky Way. Surface gravity serves as a proxy for evolutionary time, while the surface N enhancement on the y-axis takes on the role of rotational mixing efficiency. The Figure shows that N enhancement occurs earlier in the main sequence evolution (i.e., at higher log *g*) in the SMC than at higher metallicity. These early results appear to support rotational mixing which is theoretically expected to be more efficient at lower metallicity.

Regarding wind properties, Figure 5 shows the mass-loss rate of LMC and SMC supergiants studied by, respectively, Brands et al. (in preparation) and Backs et al. (in preparation) using a genetic algorithm method and the fastwind code. The results indicate that the mass loss versus luminosity relation becomes steeper in lower metallicity environments than at higher metallicity.

These early results are based on the analysis of relatively modest subsets of the XShootU data. More definitive answers will be obtained over the next few years as analyses of the full sample are completed. We expect that even basic stellar parameters such as stellar mass may need to be re-evaluated in comparison to the pre-ULLYSES era. For instance, Pauli et al. (2022) studied the earliest type eclipsing binary in the SMC, leading to a significant revision of its component masses.

## Spectral libraries and the more distant Universe

Stellar spectral libraries play a fundamental role in stellar population synthesis models, which are essential tools for studying the fundamental properties of unresolved stellar systems. Multiple empirical stellar spectral libraries, each gathered with distinct objectives, are publicly accessible. One significant limitation of the existing empirical libraries today is their coverage of hot and young stars at low metallicities.

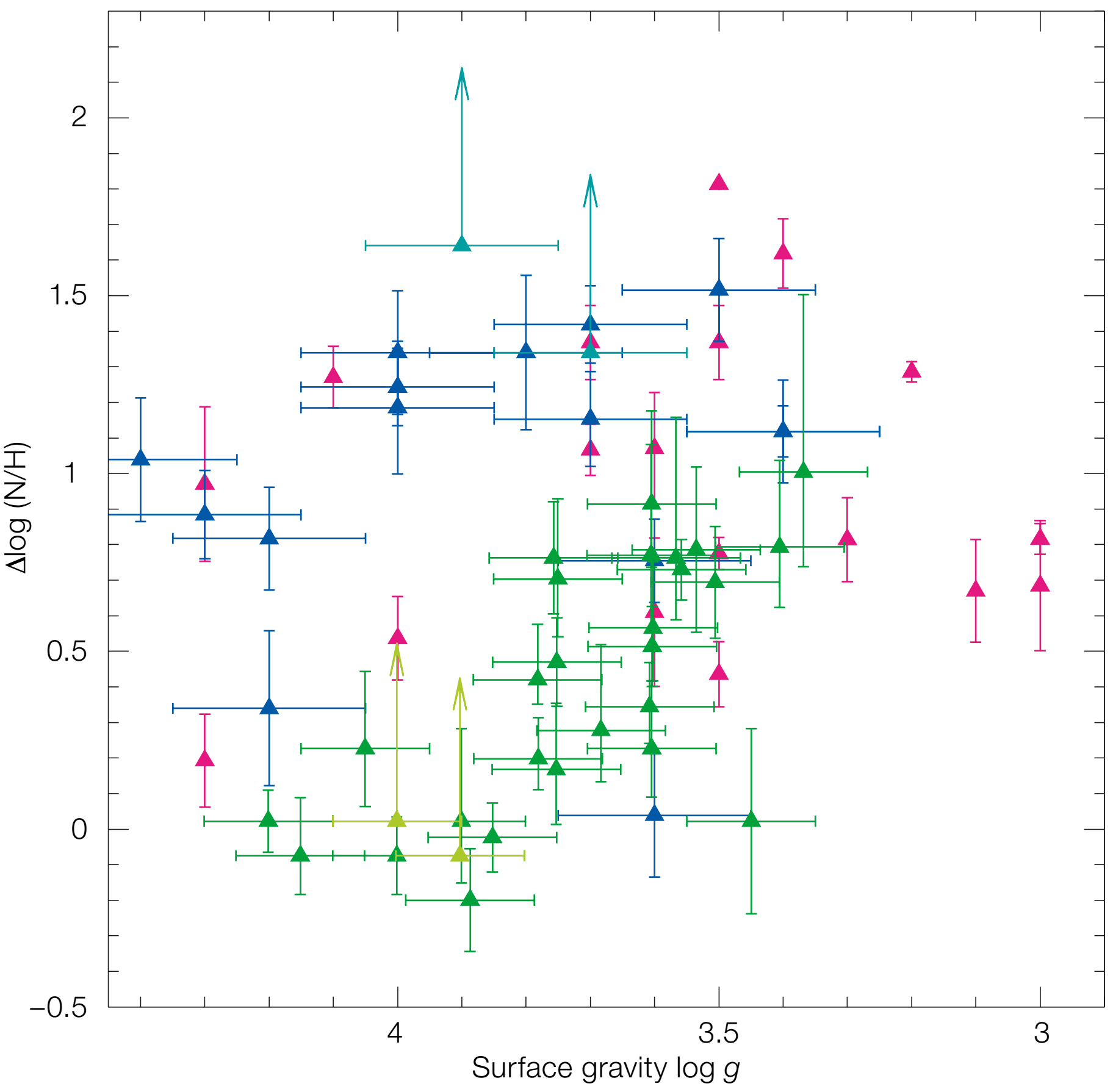

Figure 4. Logarithm of N/H minus the baseline value as a function of surface gravity for SMC/LMC/MW stars in blue/magenta/green. The light green colour is for upper or lower limits in the Galaxy, pink is for the LMC. Note that N-enrichment appears earlier in the main sequence (at higher log *g*) for the lower-*Z* environment of the SMC than in the LMC/MW. From Martins et al. (submitted to A&A).

In terms of spectral resolving power, only the X-shooter Spectral Library (XSL; Verro et al., 2022) and ELODIE archive (Moultaka et al., 2004) complement the X-Shooting ULLYSES dataset. XShootU represents the most comprehensive, highest-signal-to-noise ratio, and highest-resolution library of hot, massive stars, encompassing the broadest spectral range. Although numerous libraries cater to low-mass stars, they exhibit gaps in coverage for high-mass stars.

As an illustration, we compare the XShootU target sample with the XSL library (Verro et al., 2022), which was specifically designed for stellar population synthesis. In Figure 6, we display the Hertzsprung–Russell diagram coverage of the XSL library, revealing the absence of massive OB stars at any metallicity. XShootU excellently complements the missing parameter space of the XSL library. In tandem, these two libraries enable self-consistent population synthesis models for systems containing both young and old stars. Population synthesis models fall into two categories: semi-empirical models where the stellar evolution tracks are theoretical, but the individual stellar spectra are observed, and fully theoretical models where both components are calculated. XShootU will lead to enhancements in semi-empirical models by providing the most comprehensive spectral library of massive stars to date. Furthermore, XShootU will guide the development of new generations of atmosphere and evolution models, contributing to the refinement of fully theoretical population synthesis models.

XShootU will also provide an upgrade to the current Starburst99 (Leitherer et al., 1999) LMC + SMC library, significantly enhancing the realism of population synthesis predictions. Once this library is incorporated into Starburst99 and other population synthesis models, corresponding Cloudy photoionisation models (Ferland et al., 2017) will be computed to consider the contributions of ionised gas and dust to the integrated light of young OB star populations. These models will be made publicly available.

## Future outlook

In summary, it is anticipated that the XShootU project will furnish a wealth of data, models, and novel insights into massive stars in low-metallicity environments. It is crucial to emphasise that the overarching objective of XShootU is to establish a high-quality, consistent optical database that complements HST ULLYSES.
The resulting legacy datasets hold significant importance in ensuring the accurate interpretation of unresolved observations obtained with the JWST (Curti et al., 2023; Carnall et al., 2023). The project's subsequent aim is to deliver consistently determined stellar and wind parameters through the integration of UV and optical datasets. This phase of the of the project will not only incorporate state-of-the art non-LTE physics, but will also rigorously assess the various spectral synthesis codes and analytical methods.

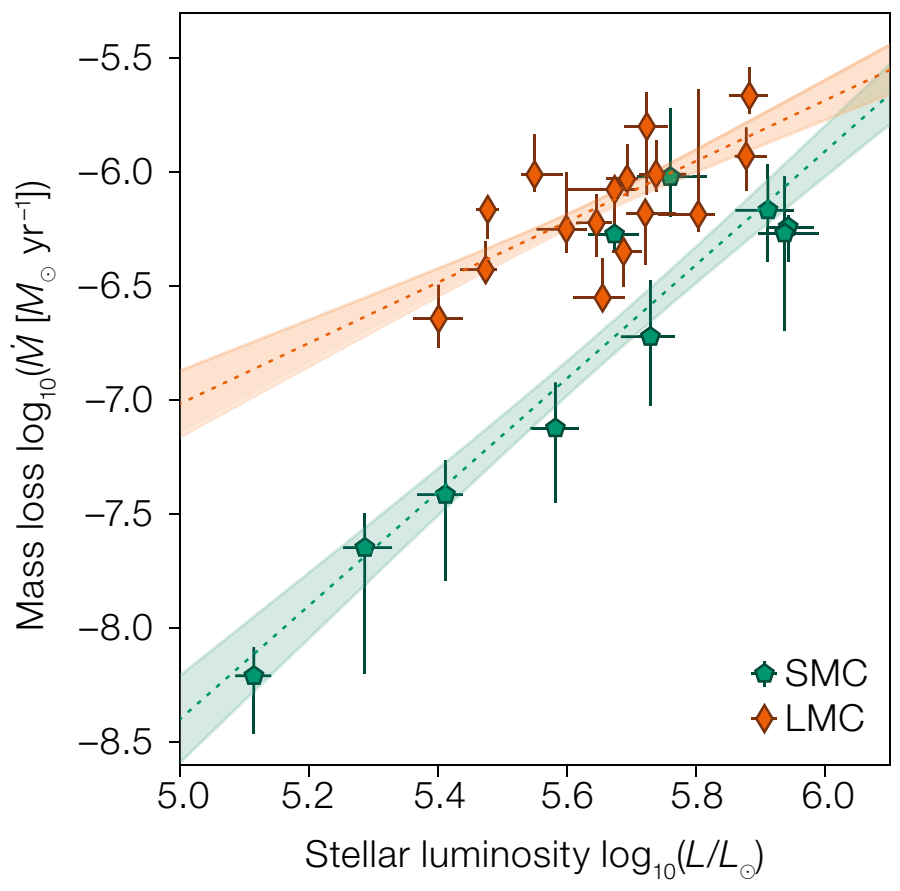

Figure 5. The empirical mass-loss luminosity relation for XShootU stars in the LMC (Brands et al., in preparation) and the SMC (Backs et al., in preparation). Note the steeper mass-loss/luminosity dependence in the SMC.

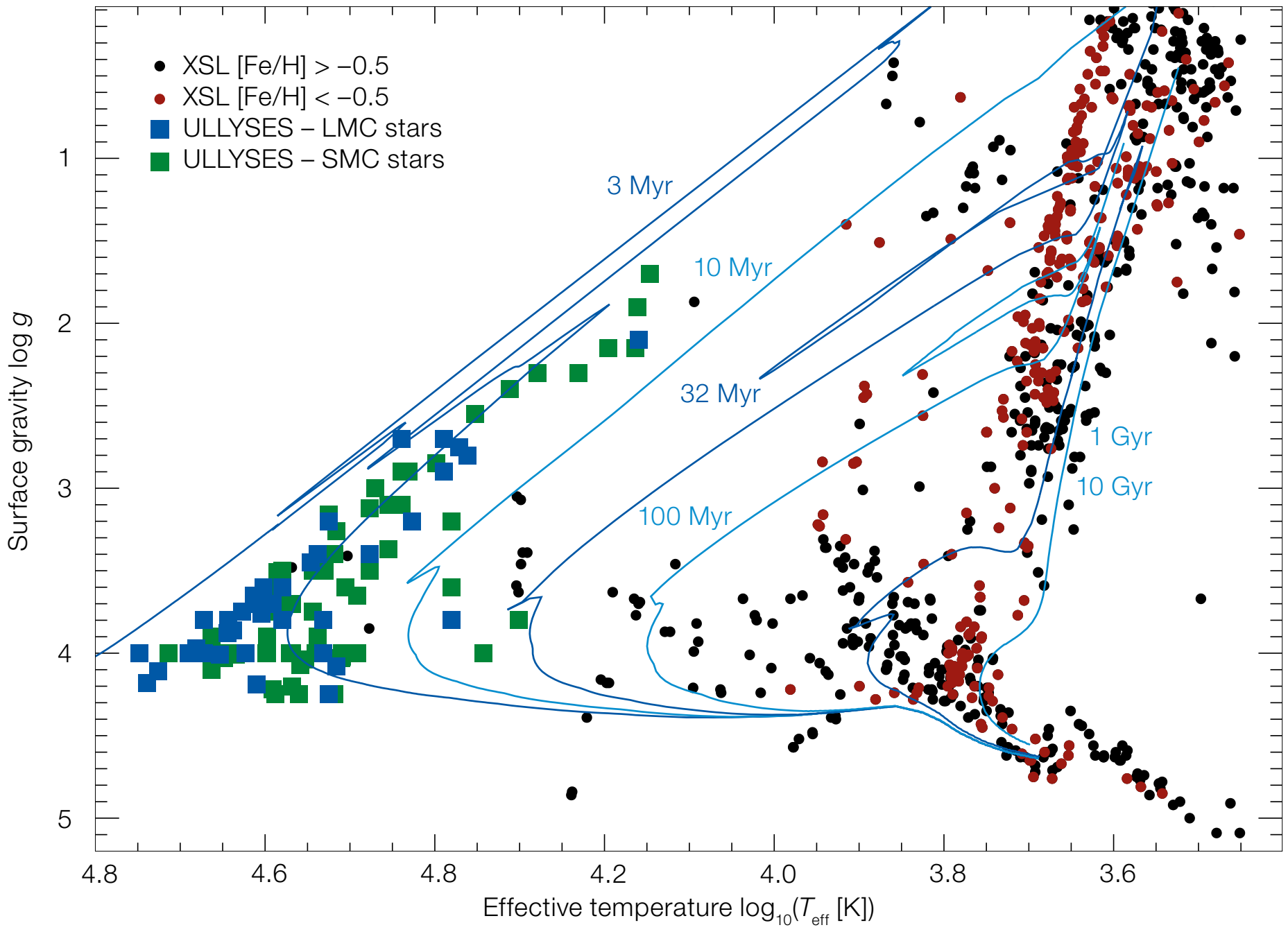

**Figure 6.** Comparison between temperatures and gravities of metal-poor and metal-rich stars in the XSL spectral library (Verro et al., 2022) and ULLYSES targets in the LMC (blue) and SMC (green).

Some of the analysis is already in progress. However, there is plenty of room for new participants to become part of this project. Additionally, the X-shooter data are accessible to the community, and we also intend to make the higher-level data products available to the broader scientific community. The high-quality XShootU data will have long-term value for a wide range of research projects, including many which may not even have been envisioned yet.

## Acknowledgements

We warmly thank the data reduction team (WG2) lead by Hugues Sana and Frank Tramper from KU Leuven for post-processing the spectra to enable quicker analysis. We also acknowledge Andrea Mehner from ESO for her help in the preparation of the OBs, and Sarah Brands and Frank Backs from the University of Amsterdam for sharing early results presented in Figure 5 on mass-loss rates versus luminosity.

## Links

[1] XShootU webpage: https://massivestars.org/xshootu/
[2] Ullyses data website: https://ullyses.stsci.edu/ullyses-download.html

ESO/Juan Carlos Muñoz Mateos

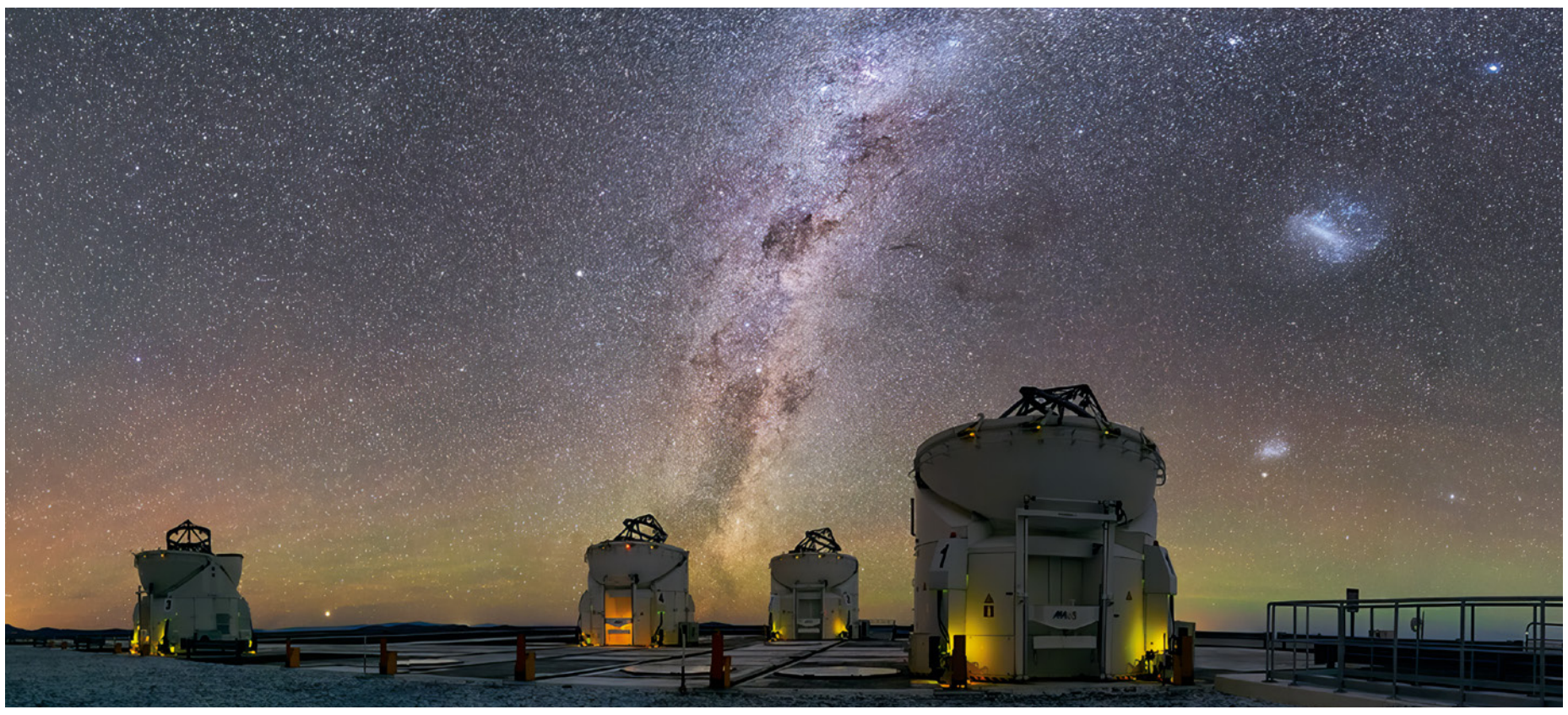

The four Auxiliary Telescopes at ESO's Paranal Observatory can be seen gazing up at the night sky in this picture. With dark and pristine skies, Paranal is one of the best places on Earth to study the Universe from. As seen in this spectacular image, the view is really full to the brim of exciting things to look at.